%% file: main.tex
\input{coml26-style.tex} 
\usepackage{amsmath, amsfonts, amssymb} 
\usepackage{graphicx}                   
\usepackage{booktabs}                   
\usepackage{array}                      
\usepackage{hyperref}                   

\begin{document}
\thispagestyle{fancy}

\title{\textbf{DMind-3}: A Sovereign Edge--Local--Cloud AI System with Controlled Deliberation and Correction-Based Tuning for Safe, Low-Latency Transaction Execution}

\authors{
  Enhao Huang\aff{1,2,3},
  Frank Li\aff{1},
  Tony Lin\aff{1},
  Lowes Yang\aff{1,*}
  
}

\affiliations{
  \affil{1}{DMind AI}
  \affil{2}{Zhejiang University}
  \affil{3}{DragonAI Technologies Ltd.}
}

{
  \renewcommand{\thefootnote}{\fnsymbol{footnote}} 
  \footnotetext[1]{Corresponding author: \texttt{team@dmind.ai}}
}

\begin{abstract}
This paper introduces DMind-3, a sovereign Edge--Local--Cloud intelligence stack designed to secure irreversible financial execution in Web3 environments against adversarial risks and strict latency constraints. While existing cloud-centric assistants compromise privacy and fail under network congestion, and purely local solutions lack global ecosystem context, DMind-3 resolves these tensions by decomposing capability into three cooperating layers: a deterministic signing-time intent firewall at the edge, a private high-fidelity reasoning engine on user hardware, and a policy-governed global context synthesizer in the cloud. We propose \textit{policy-driven selective offloading} to route computation based on privacy sensitivity and uncertainty, supported by two novel training objectives: Hierarchical Predictive Synthesis (HPS) for fusing time-varying macro signals, and Contrastive Chain-of-Correction Supervised Fine-Tuning (C$^3$-SFT) to enhance local verification reliability. Extensive evaluations demonstrate that DMind-3 achieves a 93.7\% multi-turn success rate in protocol-constrained tasks and superior domain reasoning compared to general-purpose baselines, providing a scalable framework where safety is bound to the edge execution primitive while maintaining sovereignty over sensitive user intent.
\end{abstract}

\section{Introduction}

Web3 markets have become a global financial substrate that operates continuously, clears value irreversibly, and couples risk across protocols and chains \cite{nakamoto2008bitcoin, buterin2014ethereum, wood2014ethereum}. A single user action can traverse multiple smart contracts \cite{luu2016making, chen2020survey}, trigger liquidation cascades \cite{perez2021liquidations}, or expose funds to adversarial execution through MEV \cite{daian2020flash, torres2021frontrunner}, oracle manipulation \cite{zhou2023sok}, and malicious frontends \cite{he2023txphish, su2021evil}. In this setting, decision quality is constrained by time: users often must approve a transaction in seconds while facing obfuscated calldata and rapidly moving prices \cite{eskandari2019sok}. The consequence is stark. Mistakes are not merely costly; they are final \cite{qin2021attacking}. This pressure has made AI assistance feel inevitable \cite{deepseek2025r1, wei2022chain}, yet it also exposes a mismatch between what current AI systems provide and what Web3 execution demands.

Most deployed assistants follow a cloud-centric pattern \cite{chen2019deep}. They can be capable at explanation and summarization \cite{yang2025qwen3, vaswani2017attention}, but they are poorly placed at the moment that matters most: the signature. Latency becomes unpredictable under network congestion \cite{mao2017survey}, precisely when volatility rises and users need faster, not slower, guidance. Privacy is fragile because portfolios, intents, and strategies are routinely transmitted to remote services \cite{kairouz2021advances, bonawitz2017practical}. Robustness is weaker than it appears because advice is not enforcement, and a helpful explanation does not prevent an unsafe approval \cite{huang2024user, erinle2025sok}. When adversaries can influence inputs through spoofed interfaces, misleading prompts, or carefully constructed calldata \cite{liu2023prompt, debenedetti2024agentdojo}, a cloud assistant that only responds in text cannot reliably bind its guidance to the actual action being authorized.

Pushing intelligence to the edge seems like an obvious fix \cite{shi2016edge, kang2017neurosurgeon}, but a purely on-device solution is also incomplete. The signing boundary is the right place to enforce safety \cite{sabt2015trusted, arnautov2016scone}, yet thin clients cannot maintain the global context that makes many Web3 risks legible. The safety of a router depends on recent ecosystem behavior \cite{wang2022cyclic}; the risk of a lending position depends on cross-market correlations \cite{werner2022sok}; the meaning of a governance action depends on broader policy and macro conditions. At the same time, the most sensitive context, including private portfolio state and proprietary strategies, should not be exported by default \cite{evans2018pragmatic, knott2021crypten}. These constraints suggest that the problem is not choosing between local and cloud intelligence. The problem is designing an end-to-end system that places computation across edge, local, and cloud tiers with explicit trust boundaries, predictable performance, and enforceable behavior at the moment of execution \cite{hunt2018ryoan}.

This paper presents \textbf{DMind-3}, a sovereign Edge--Local--Cloud intelligence stack designed for adversarial financial execution in Web3. DMind-3 decomposes capability into three cooperating layers. At the edge, a lightweight component runs inside the browser or wallet to translate raw transaction payloads into user-facing intent and to enforce signing-time policies that can block, throttle, or require extra verification before a signature is produced \cite{tsankov2018securify, tolmach2021survey}. On a user-controlled machine, a higher-fidelity component performs deeper reasoning tasks such as contract interpretation, economic exploit analysis, and strategy simulation while keeping private state local \cite{yao2023react, deepseek2024v3}. In the cloud, a scalable service synthesizes macro context from ecosystem-wide signals and coordinates tools and agent workflows when policy permits, without assuming access to private intent \cite{mcmahan2017communication, hinton2015distilling}. The system is structured so that the final decision and enforcement bind to the transaction bytes at the edge, while heavier inference and broader context are selectively incorporated through policy-governed offloading \cite{cheng2017survey, gou2021knowledge}.

\begin{figure}[t!]
    \centering
    \includegraphics[width=0.9\linewidth]{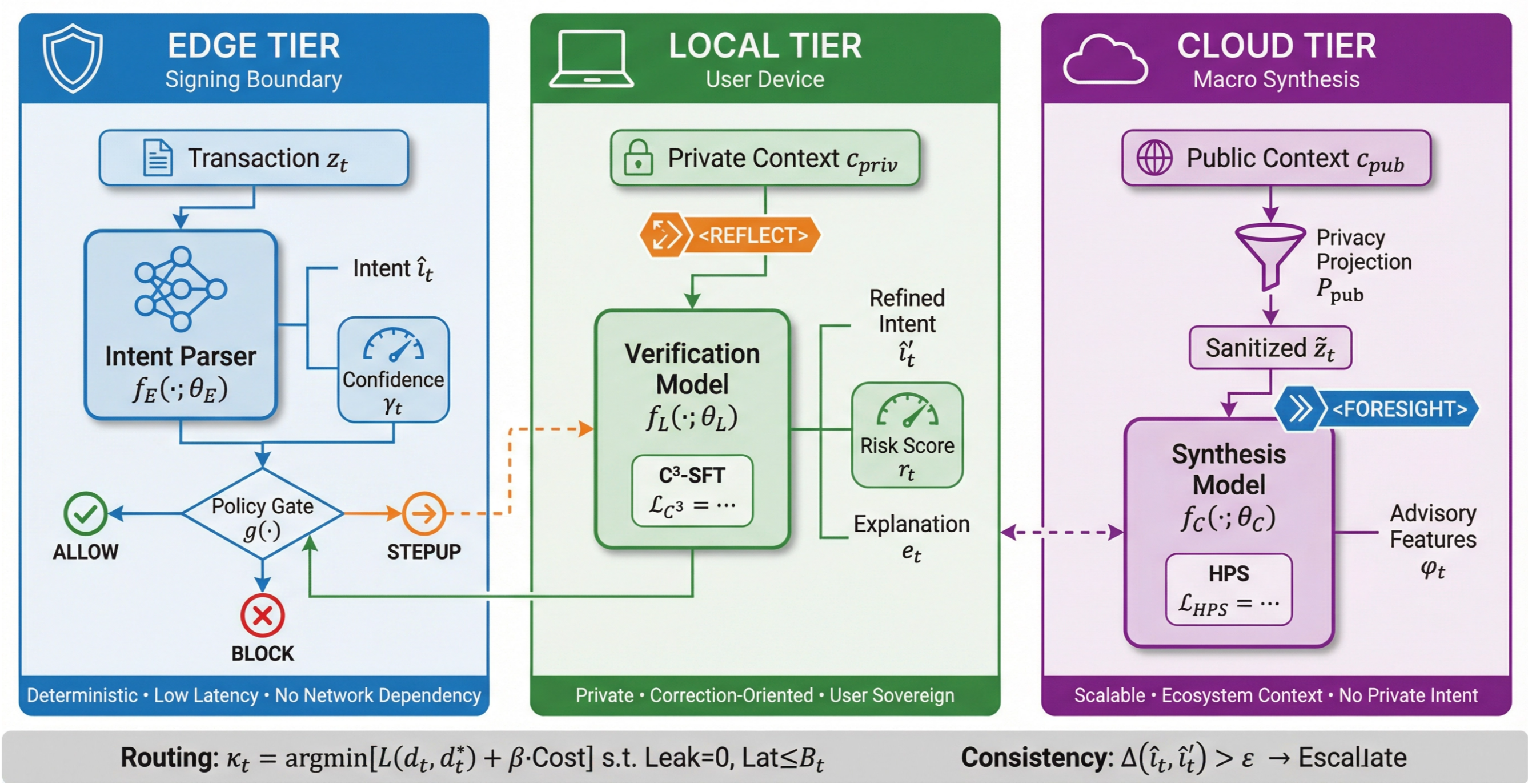}
    \caption{The DMind-3 Edge--Local--Cloud Architecture: A sovereign stack for Web3 execution. The system routes transactions based on privacy, latency, and uncertainty, ensuring that final enforcement (Policy Gate) happens at the deterministic Edge Tier.}
    \label{fig:architecture}
\end{figure}

Designing such a stack requires resolving three systems tensions. First, safety must be anchored to the signature, not to explanation. DMind-3 treats the edge layer as an intent firewall that parses calldata, identifies high-risk patterns such as unlimited approvals or unusual delegate behavior, and applies deterministic rules that do not depend on network availability \cite{zhang2019security, conti2018survey}. Second, global context is valuable, but sovereignty requires that sensitive context remains confined. DMind-3 introduces policy-driven selective offloading that classifies inputs by sensitivity, latency criticality, and uncertainty, then routes subproblems to the appropriate tier \cite{warden2019tinyml}. The cloud can propose hypotheses and ecosystem-level risk context \cite{weintraub2022flash}, while the local tier binds recommendations to private state and the edge tier enforces the final action. Third, the system must remain robust under adversarial inputs and volatile conditions \cite{goodfellow2014explaining, carlini2017towards, madry2018towards}. DMind-3 incorporates a risk-aware orchestration plane that tracks provenance of signals and tools, checks consistency across tiers, and escalates to stronger verification modes when uncertainty rises \cite{wang2019neural, gu2017badnets}, enabling graceful degradation under poor connectivity or partial service failure. The architectural flow and the routing logic governing these transitions are detailed in Figure~\ref{fig:architecture}.

DMind-3 operationalizes a broader principle: sovereignty is an end-to-end property of the intelligence pipeline \cite{feng2024survey, chen2023verified}. It is not achieved by simply running a model locally, nor by trusting a remote service to behave well. Sovereignty requires a disciplined separation of roles, explicit trust contracts between tiers, and an enforcement mechanism that is inseparable from execution \cite{wang2023mixers, tolmach2021formal}. While this work is motivated by Web3, the architectural lesson generalizes to other domains where decisions are time-critical, irreversible, and adversarial \cite{schwarz2022three, ren2019analysis}.

\textbf{Contributions.} This paper makes four contributions. First, we present an Edge--Local--Cloud architecture for sovereign intelligence in adversarial financial execution, including a signing-time intent firewall at the edge, a private high-fidelity reasoning tier on user hardware, and a macro-context synthesis tier in the cloud. Second, we introduce policy-driven selective offloading that routes computation by privacy sensitivity, latency criticality, and uncertainty, enabling the system to exploit global context without exporting private intent by default. Third, we design a risk-aware orchestration plane that performs cross-tier consistency checks, provenance tracking, and uncertainty-triggered escalation to stronger verification modes. Fourth, we develop an end-to-end evaluation methodology for sovereign intelligence stacks in Web3, measuring signing-time latency and tail behavior under network degradation, privacy leakage under realistic workloads, and robustness against transaction obfuscation and interface-level attacks.

\section{Related Work}

\subsection{Security in Decentralized Systems}

The security of decentralized systems, particularly blockchains, has been a central theme of academic research since the inception of Bitcoin \cite{nakamoto2008bitcoin}. Early work focused on the fundamental security properties of consensus protocols, such as Nakamoto consensus \cite{ren2019analysis}, and their resilience against attacks \cite{conti2018survey, schwarz2022three}. With the advent of Ethereum \cite{buterin2014ethereum, wood2014ethereum} and its support for smart contracts, the research landscape expanded dramatically to address the unique vulnerabilities introduced by programmable on-chain logic.

A significant body of work has focused on identifying and mitigating vulnerabilities within smart contracts. Pioneering research by Luu et al. \cite{luu2016making} identified several critical vulnerabilities, including transaction-ordering dependence and reentrancy, the latter of which was famously exploited in the DAO hack. This spurred the development of numerous static and dynamic analysis tools, such as Oyente \cite{luu2016making} and Securify \cite{tsankov2018securify}, designed to automatically detect such flaws. Comprehensive surveys by Chen et al. \cite{chen2020survey} and Tolmach et al. \cite{tolmach2021survey} provide a systematic overview of the vulnerabilities, attacks, and formal verification techniques in the Ethereum ecosystem. Zhang et al. \cite{zhang2019security} further examined the broader security and privacy landscape of blockchain technology.

More recently, the concept of Miner Extractable Value (MEV), later generalized to Maximal Extractable Value, has become a dominant topic. The seminal work by Daian et al., ``Flash Boys 2.0'' \cite{daian2020flash}, exposed the ``dark forest'' of transaction frontrunning and arbitrage bots operating on decentralized exchanges (DEXs). Torres et al. \cite{torres2021frontrunner} conducted a large-scale empirical study quantifying the prevalence and profitability of frontrunning attacks, while Eskandari et al. \cite{eskandari2019sok} systematized knowledge on transparent dishonesty and front-running. The rise of complex DeFi protocols \cite{werner2022sok} has also introduced novel attack vectors, such as flash loan attacks \cite{qin2021attacking} and liquidation cascades \cite{perez2021liquidations}, which are systematically analyzed by Zhou et al. \cite{zhou2023sok}. Wang et al. \cite{wang2022cyclic} studied cyclic arbitrage patterns in DEXs, while Weintraub et al. \cite{weintraub2022flash} measured MEV extraction in private transaction pools like Flashbots. The formal analysis of these composable DeFi protocols is an active area of research, aiming to ensure their security and stability \cite{tolmach2021formal}. Our work, DMind-3, contributes to this line of research by proposing a proactive defense mechanism against MEV-related threats at the edge, before transactions are broadcast to the public mempool.

\subsection{Intelligence at the Edge and in Decentralized Environments}

The integration of artificial intelligence with edge computing has emerged as a powerful paradigm for delivering low-latency, privacy-preserving intelligent services \cite{shi2016edge}. The survey by Chen and Ran \cite{chen2019deep} provides a comprehensive review of the state-of-the-art at the intersection of deep learning and edge computing, highlighting applications and system design challenges. Mao et al. \cite{mao2017survey} examined mobile edge computing from the communication perspective, addressing the fundamental tradeoffs between computation offloading and communication overhead. Kang et al. \cite{kang2017neurosurgeon} proposed Neurosurgeon, a framework for collaborative intelligence between cloud and mobile edge that dynamically partitions DNN computation.

A key challenge in this domain is the efficient deployment of large AI models on resource-constrained edge devices. This has led to extensive research in model compression techniques, including quantization, pruning, and low-rank factorization, as surveyed by Cheng et al. \cite{cheng2017survey}. Knowledge distillation, introduced by Hinton et al. \cite{hinton2015distilling}, offers a powerful method for transferring knowledge from a large ``teacher'' model to a smaller ``student'' model, which is particularly relevant for edge deployment. Gou et al. \cite{gou2021knowledge} provided a comprehensive survey of knowledge distillation techniques and their applications. The field of TinyML further pushes the boundaries of on-device machine learning, enabling inference on ultra-low-power microcontrollers \cite{warden2019tinyml}.

In parallel, the field of federated learning, pioneered by McMahan et al. \cite{mcmahan2017communication}, has provided a framework for training machine learning models on decentralized data without compromising user privacy. Kairouz et al. \cite{kairouz2021advances} systematically documented advances and open problems in federated learning, while Bonawitz et al. \cite{bonawitz2017practical} developed practical secure aggregation protocols for privacy-preserving machine learning. This approach, where model updates are trained locally on devices and then aggregated centrally, aligns closely with the decentralized ethos of Web3. DMind-3 builds upon these foundations by creating a hierarchical architecture where a global model, analogous to a central federated learning server, coordinates a network of specialized, compressed models deployed at the edge. These edge models perform localized, real-time analysis of transaction risks, effectively creating a decentralized, intelligent pre-consensus layer.

\subsection{Large Language Models and AI Agent Security}

The rapid advancement of large language models (LLMs) has fundamentally transformed the landscape of AI capabilities. The Transformer architecture \cite{vaswani2017attention} laid the foundation for modern LLMs, enabling unprecedented performance in natural language understanding and generation. The year 2024-2025 witnessed remarkable progress in frontier models. DeepSeek-V3 \cite{deepseek2024v3} introduced a 671B-parameter Mixture-of-Experts architecture with only 37B activated parameters per token, achieving performance comparable to leading closed-source models at a fraction of the training cost. DeepSeek-R1 \cite{deepseek2025r1}, published in Nature, demonstrated that reasoning capabilities can be incentivized through pure reinforcement learning without human-labeled reasoning trajectories, exhibiting emergent patterns such as self-reflection and dynamic strategy adaptation. The subsequent DeepSeek-V3.2 \cite{deepseek2025v32} introduced sparse attention mechanisms and achieved gold-medal performance at the 2025 International Mathematical Olympiad. Alibaba's Qwen3 \cite{yang2025qwen3} unified thinking and non-thinking modes into a single framework, supporting 119 languages with parameter scales from 0.6B to 235B. Google's Gemini 2.5 \cite{gemini2025report} pushed the frontier in multimodal understanding with the ability to process up to 3 hours of video content. Anthropic's Claude 4 \cite{anthropic2025claude4} and OpenAI's GPT-5 \cite{openai2025gpt5} further advanced the state of the art in enterprise AI applications.

These models have enabled the creation of sophisticated AI agents capable of multi-step reasoning through techniques such as chain-of-thought prompting \cite{wei2022chain} and the ReAct framework \cite{yao2023react}, which synergizes reasoning and acting to solve complex tasks. Wang et al. \cite{wang2024survey} provided a comprehensive survey of LLM-based autonomous agents, while Plaat et al. \cite{plaat2024reasoning} systematically reviewed reasoning capabilities in large language models.

However, as these agents become more autonomous and are granted access to external tools and APIs, their security becomes a paramount concern. A primary threat is prompt injection, where an attacker can manipulate the agent's instructions by embedding malicious commands in the data it processes \cite{liu2023prompt}. This can lead to the agent performing unintended and potentially harmful actions. Recent work has also revealed that even fine-tuning aligned language models can compromise their safety \cite{qi2024finetuning}. To address these emerging threats, researchers have developed comprehensive benchmarks and evaluation frameworks like AgentDojo \cite{debenedetti2024agentdojo} and Agent Security Bench \cite{zhang2024agent} to formalize and assess the security of LLM agents. Chu et al. \cite{chu2024jailbreak} provided a comprehensive assessment of jailbreak attacks against LLMs. The challenge of hallucination detection has also received significant attention, with Farquhar et al. \cite{farquhar2024detecting} proposing semantic entropy-based methods published in Nature, and Du et al. \cite{du2024haloscope} introducing HaloScope for leveraging unlabeled generations.

Furthermore, the broader topics of AI safety and alignment, including adversarial attacks \cite{goodfellow2014explaining, carlini2017towards, madry2018towards} and backdoor attacks \cite{wang2019neural, gu2017badnets}, are highly relevant. The principles of secure multi-party computation (MPC) \cite{evans2018pragmatic, knott2021crypten} and confidential computing \cite{feng2024survey, chen2023verified} also offer valuable paradigms for building trustworthy distributed AI systems. Trusted execution environments (TEEs) \cite{sabt2015trusted, arnautov2016scone, hunt2018ryoan} provide hardware-based isolation for sensitive computations.

In the Web3 context, recent work has examined user-perceived security risks \cite{huang2024user} and cryptocurrency wallet vulnerabilities \cite{erinle2025sok}. He et al. \cite{he2023txphish} studied transaction simulation-based phishing attacks, while Su et al. \cite{su2021evil} discovered attacks on Ethereum decentralized applications. Wang et al. \cite{wang2023mixers} analyzed how zero-knowledge proof blockchain mixers affect user privacy. DMind-3 directly addresses the agent security problem by incorporating a trust-aware execution environment. The central LLM agent in our architecture operates with a principle of least privilege, and its interactions with the edge models and the blockchain are mediated by security policies and formal verification, ensuring that the agent's actions remain aligned with the system's security goals.

\section{Theory and Methods}
\label{sec:theory_methods}

DMind-3 targets a specific operational regime: a user must authorize an irreversible transaction under tight latency, imperfect information, and adversarial pressure. The system goal is therefore not to maximize conversational helpfulness, but to minimize \emph{decision-critical errors at signing time} while preserving sovereignty over sensitive intent. DMind-3 achieves this by co-designing (i) an Edge-Local-Cloud (ELC) execution topology with explicit trust boundaries and (ii) training objectives that favor correction-oriented verification and risk-aware synthesis. This section formalizes the ELC model, the routing and escalation logic, and the two method components used during model development: Hierarchical Predictive Synthesis (HPS) and Contrastive Chain-of-Correction Supervised Fine-Tuning (C\textsuperscript{3}-SFT). The core objectives and inference rules match the model card.

\subsection{Setting and Edge-Local-Cloud System Model}
We consider a stream of transaction requests indexed by $t \in \{1,\dots,T\}$. Each request exposes a transaction payload $z_t$ (destination, calldata, value transfer, approvals, and parameters) that fully determines what will execute. The system also has access to user-private context $c^{\mathrm{priv}}_t$ (preferences, allowlists, exposure limits, local history) and policy-permitted public context $c^{\mathrm{pub}}_t$ (non-sensitive metadata and aggregated signals). DMind-3 produces (i) a structured intent interpretation $\hat{\imath}_t$, (ii) a signing decision $d_t \in \{\textsc{Allow}, \textsc{Block}, \textsc{StepUp}\}$, and (iii) an explanation $e_t$. Explanations improve usability, but DMind-3 is designed such that the safety outcome does not depend on the user reading text under time pressure.

The ELC topology assigns distinct responsibilities to each tier. The \emph{Edge} tier $E$ runs at the signing boundary and binds safety to the payload bytes. The \emph{Local} tier $L$ runs on user-controlled hardware and performs higher-fidelity verification using private context. The \emph{Cloud} tier $C$ provides scalable synthesis and orchestration when policy permits, without assuming access to private intent. Each tier hosts a model (or ensemble) parameterized by $\theta_E, \theta_L, \theta_C$.

At the edge, DMind-3 extracts intent from the payload and computes a confidence score:
\begin{equation}
\hat{\imath}_t = f_E(z_t;\theta_E),
\qquad
\gamma_t = \mathrm{conf}_E(\hat{\imath}_t).
\label{eq:edge_intent}
\end{equation}
The edge tier then applies a local policy $\Pi$ through a deterministic decision gate:
\begin{equation}
d_t = g(\hat{\imath}_t,\Pi,\gamma_t).
\label{eq:edge_gate}
\end{equation}
A key design choice is conservatism under ambiguity: if intent cannot be interpreted with sufficient confidence, the default is $\textsc{StepUp}$ rather than $\textsc{Allow}$. This yields a safety property that is resilient to partial connectivity, because the decision gate depends only on the payload being signed and local policy state.

When $d_t=\textsc{StepUp}$, DMind-3 routes the request to the local tier for deeper verification:
\begin{equation}
(\hat{\imath}'_t, r_t, e_t) = f_L(z_t, c^{\mathrm{priv}}_t;\theta_L),
\label{eq:local_verify}
\end{equation}
where $\hat{\imath}'_t$ refines the intent interpretation and $r_t$ is a risk score that can incorporate user-specific constraints. When policy allows, the cloud tier produces non-sensitive synthesis features from a sanitized projection of the request:
\begin{equation}
\tilde{z}_t = \mathcal{P}_{\mathrm{pub}}(z_t,\Pi),
\qquad
\phi_t = f_C(\tilde{z}_t, c^{\mathrm{pub}}_t;\theta_C).
\label{eq:cloud_features}
\end{equation}
Here $\mathcal{P}_{\mathrm{pub}}$ removes or coarsens sensitive attributes under $\Pi$. In DMind-3, $\phi_t$ is treated as advisory context that can improve routing and verification depth, while the signing-time enforcement remains anchored at the edge gate in (\ref{eq:edge_gate}).

\subsection{Policy-Constrained Routing and Risk Escalation}
The systems question is how to place computation while meeting privacy and latency constraints. DMind-3 models routing as selecting a compute plan $\kappa_t$ from a small set $\mathcal{K}$ (for example, edge-only, edge-to-local, edge-to-cloud, edge-to-cloud-to-local). We write routing as a constrained objective that trades expected decision loss against runtime cost:
\begin{equation}
\kappa_t
=
\arg\min_{\kappa \in \mathcal{K}}
\;
\mathbb{E}\!\left[\ell(d_t(\kappa), d_t^\star)\right]
+
\beta \cdot \mathrm{Cost}(\kappa)
\quad
\text{s.t.}
\quad
\mathrm{Leak}(\kappa,\Pi)=0,\;
\mathrm{Lat}(\kappa) \le B_t.
\label{eq:routing_obj}
\end{equation}
Here $d_t^\star$ denotes an idealized safe decision under complete information, $\ell(\cdot)$ penalizes unsafe authorizations more heavily than conservative blocking, $\mathrm{Cost}(\kappa)$ captures compute and tool overhead, and $\mathrm{Lat}(\kappa)$ is predicted latency under current conditions with budget $B_t$. The constraint $\mathrm{Leak}(\kappa,\Pi)=0$ encodes the sovereignty requirement that sensitive intent is not exported to the cloud by default. In implementation, DMind-3 uses lightweight predictors and policy rules to approximate (\ref{eq:routing_obj}), avoiding claims of globally optimal routing while retaining a clear formal target.

Escalation is triggered by uncertainty and lightweight risk patterns observable at the edge. A representative decision rule is:
\begin{equation}
\textsc{StepUp}
\;\;\text{if}\;\;
\big(\gamma_t < \tau_{\mathrm{conf}}\big)
\;\;\vee\;\;
\big(\rho_E(\hat{\imath}_t) > \tau_{\mathrm{risk}}\big),
\label{eq:stepup_rule}
\end{equation}
where $\rho_E(\cdot)$ is a compact risk heuristic over intent features (for example, permission amplification patterns or unusual approval structure). The thresholds $\tau_{\mathrm{conf}}$ and $\tau_{\mathrm{risk}}$ are policy defaults that can be tuned without changing model weights. To mitigate adversarial UI mismatch, DMind-3 also checks cross-tier semantic agreement. Let $\Delta(\hat{\imath}_t,\hat{\imath}'_t)$ measure divergence between edge and local intent interpretations; DMind-3 tightens verification when:
\begin{equation}
\Delta(\hat{\imath}_t,\hat{\imath}'_t) > \epsilon.
\label{eq:consistency}
\end{equation}
This mechanism is intentionally simple: it does not require a perfect detector, only a reliable trigger that reallocates budget from speed to assurance when inconsistency appears.

\subsection{Controlled Deliberation via Dual-State Inference}
Transaction-time systems need a predictable way to switch between a fast path and a verification path. DMind-3 implements this as controlled deliberation: a dual-state inference mechanism activated by a trigger token $\tau$. In the cloud tier, the model card formulation is:
\begin{equation}
\hat{y} = 
\begin{cases} 
\operatorname*{arg\,max}\limits_{y} P_\theta(y \mid x, \mathcal{C}_{\text{global}}) & \text{if } \tau = \emptyset \quad (\text{Standard Mode}) \\
\operatorname*{arg\,max}\limits_{y} P_\theta(y \mid x, \mathcal{C}_{\text{global}}, \mathcal{R}_{\text{risk}}, \mathcal{H}_{\text{hist}}) & \text{if } \tau = \texttt{<FORESIGHT>} \quad (\text{Strategic Mode})
\end{cases}
\label{eq:oracle_dual}
\end{equation}
This does not assert perfect prediction; rather, it provides a disciplined interface for risk-aware synthesis when the system deems macro context helpful.

In the local tier, the model card formulation is:
\begin{equation}
\hat{y} = 
\begin{cases} 
\operatorname*{arg\,max}\limits_{y} P_\theta(y \mid x) & \text{if } \tau = \emptyset \quad (\text{Standard Mode}) \\
\operatorname*{arg\,max}\limits_{y} P_\theta(y \mid x, \mathcal{G}_{neg}(x), \mathcal{H}_{crit}) & \text{if } \tau = \texttt{<REFLECT>} \quad (\text{Audit Mode})
\end{cases}
\label{eq:local_dual}
\end{equation}
where $\mathcal{G}_{neg}(x)$ denotes a latent negative hypothesis and $\mathcal{H}_{crit}$ is a critique operator that emphasizes verification. The key property is measurability: verification is a controlled state transition tied to system triggers such as (\ref{eq:stepup_rule}) and (\ref{eq:consistency}), rather than an ad hoc increase in prompt verbosity.

\subsection{Training Objectives: HPS and C\textsuperscript{3}-SFT}
DMind-3 uses two complementary training objectives aligned with the roles of the cloud and local tiers. The cloud tier employs Hierarchical Predictive Synthesis (HPS) to fuse time-varying, potentially conflicting signals into a probabilistic view of future states. The HPS objective from the model card is:
\begin{equation}
\mathcal{L}_{\text{HPS}}(\theta) = - \mathbb{E}_{\mathcal{D}} \left[ \sum_{t=1}^{T} \sum_{i=1}^{M} \omega_i \cdot \log P_\theta(S'_{t+1} \mid S_t, A_t, M_i) \right] + \lambda \sum_{l=1}^{L} \| \Omega_l(\theta) - \Omega_l(\theta_{\text{ref}}) \|_F.
\label{eq:hps_loss}
\end{equation}
Here $S_t$ denotes the global state at time $t$, $A_t$ the set of actions at time $t$, and $M_i$ the $i$th modality; $\omega_i$ are modality importance weights. The regularization term stabilizes training relative to a reference model $\theta_{\text{ref}}$ through layer-wise parameter matrices $\Omega_l(\cdot)$. In the system, HPS is primarily used to produce synthesis features $\phi_t$ in (\ref{eq:cloud_features}) that improve routing and verification depth without relying on any single signal stream. We emphasize the operational role of HPS as calibrated synthesis for decision support, rather than as an absolute claim of foresight.

The local tier is trained to reduce a different failure mode: fluent but incorrect reasoning that changes signing decisions. DMind-3 uses Contrastive Chain-of-Correction Supervised Fine-Tuning (C\textsuperscript{3}-SFT) to train the model to traverse a correction trajectory by contrasting against plausible negative reasoning. The model card objective is:
\begin{equation}
\mathcal{L}_{C^3}(\theta) = - \mathbb{E}_{\mathcal{D}} \left[ \sum_{t=1}^{T} \alpha_t \cdot \log P_\theta(y^+_{t} \mid x, y^-, y^+_{<t}) \right] + \lambda \mathbb{KL}[\pi_\theta || \pi_{\text{ref}}],
\label{eq:c3_loss}
\end{equation}
where $\mathcal{D} = \{(x, y^-, y^+_{cot})\}_{i=1}^N$ is the triplet dataset, $y^-$ is a negative sample containing common logical fallacies, and $y^+_{cot}$ is the corrective trajectory; $\alpha_t$ weights decision-critical correction steps. The KL term regularizes the fine-tuned policy $\pi_\theta$ toward a reference $\pi_{\text{ref}}$ to preserve general capabilities while specializing for audit behavior. In deployment, the C\textsuperscript{3}-SFT trained model is invoked predominantly in the \texttt{<REFLECT>} state in (\ref{eq:local_dual}), matching the system intent: use the fast path when confidence is high, and allocate heavier reasoning when risk or ambiguity warrants it.

Taken together, the ELC trust contract determines where decisions are made, the routing objective in (\ref{eq:routing_obj}) determines when computation is escalated, the dual-state inference rules in (\ref{eq:oracle_dual}) and (\ref{eq:local_dual}) determine how deliberation is controlled, and the training objectives in (\ref{eq:hps_loss}) and (\ref{eq:c3_loss}) shape the respective tiers toward synthesis and correction. This integration yields a system that is fast by default, more deliberate under uncertainty, and sovereign by construction, while avoiding brittle assumptions about constant connectivity or sustained user attention.

\section{Experimental Evaluation}
\label{sec:eval}

We evaluate DMind-3 in the setting that motivates its design: transaction execution under time pressure, imperfect information, and adversarially crafted inputs. In this regime, strong general reasoning is necessary, but reliability hinges on two additional properties: the edge component must behave like a protocol-faithful execution primitive, and the overall stack must deliver predictable latency while keeping sensitive intent on user-controlled hardware by default. Our evaluation therefore combines model capability benchmarks with transaction-time structured behavior tests, and then adds a small set of system checks that are easy to reproduce with logging and instrumentation.

\subsection{Models, Baselines, and Protocol}
We report results for three models in the DMind-3 stack: \textbf{DMind-3} (21B, cloud-tier oracle), \textbf{DMind-3-mini} (4B, local-tier verifier), and \textbf{DMind-3-nano} (edge-tier signer-side assistant). We compare DMind-3 and DMind-3-mini against representative baselines commonly used in practice: GPT-5.1, Claude Sonnet 4.5, DeepSeek V3.2, and MiniMax M2.1. For edge-tier structured behavior, we compare DMind-3-nano to functiongemma-270m-it and Qwen3-0.6B, which are compact models frequently used for tool and protocol tasks.

Unless otherwise stated, scores follow each benchmark's standard convention (higher is better). For system microbenchmarks, we report latency percentiles measured from end-to-end instrumentation of the signing path, including parsing, policy checks, and (when applicable) escalation to local verification and sanitized cloud synthesis.

\subsection{Backbone Capability on Knowledge and Reasoning Benchmarks}
We first verify that the cloud and local backbones provide competitive competence on domain and reasoning benchmarks. Table~\ref{tab:backbone_scores} summarizes results on three evaluations: \textbf{DMind Benchmark} (domain knowledge and reasoning for Web3 transaction contexts), \textbf{FinanceQA} (financial QA), and \textbf{AIME 2025} (advanced mathematical reasoning).

DMind-3 leads the DMind Benchmark and FinanceQA while remaining near parity on AIME 2025. This pattern is consistent with the intended division of labor: the cloud tier should maintain strong general reasoning while being particularly reliable in transaction-relevant domains. DMind-3-mini, despite its smaller footprint, retains much of the domain performance needed for local verification, which is valuable because local verification is invoked under uncertainty and must remain capable without requiring a cloud round trip.

\begin{table}[t]
\centering
\small
\setlength{\tabcolsep}{6pt}
\begin{tabular}{lccc}
\toprule
\textbf{Model} & \textbf{DMind Benchmark} & \textbf{FinanceQA} & \textbf{AIME 2025} \\
\midrule
DMind-3 (21B) & 80.3 & 70.3 & 93.3 \\
GPT-5.1 & 78.1 & 68.4 & 94.0 \\
Claude Sonnet 4.5 & 76.9 & 63.1 & 87.0 \\
DMind-3-mini (4B) & 76.2 & 65.2 & 83.3 \\
DeepSeek V3.2 & 70.1 & 67.8 & 89.3 \\
MiniMax M2.1 & 66.1 & 66.1 & 81.0 \\
\bottomrule
\end{tabular}
\caption{Backbone capability on three benchmarks. Higher is better under each benchmark's scoring protocol.}
\label{tab:backbone_scores}
\end{table}

\subsection{Transaction-Time Structured Behavior at the Signing Boundary}
Backbone capability does not guarantee safe execution behavior. At signing time, the edge tier must follow strict constraints: select the correct action primitive, extract arguments accurately, adhere to protocol formats, and complete multi-step interactions that resemble real approval workflows. We evaluate DMind-3-nano on a structured interaction benchmark that measures four metrics: \textbf{function recognition}, \textbf{argument extraction}, \textbf{protocol adherence}, and \textbf{multi-turn success}. Table~\ref{tab:nano_scores} shows that DMind-3-nano performs strongly on all four.

The most operationally important metric is multi-turn success. Transaction verification often requires a second pass: resolving ambiguity, re-checking intent, or stepping up to stricter checks after a suspicious pattern. DMind-3-nano achieves 93.7 multi-turn success, which is substantially higher than the compact baselines. This supports the design choice to treat the edge tier as a constrained, protocol-faithful component that can safely sit on the critical path.

\begin{table}[t]
\centering
\small
\setlength{\tabcolsep}{6pt}
\begin{tabular}{lcccc}
\toprule
\textbf{Model} & \textbf{Func. Rec.} & \textbf{Arg. Ext.} & \textbf{Prot. Adh.} & \textbf{Multi-turn} \\
\midrule
DMind-3-nano & 98.8 & 97.4 & 97.9 & 93.7 \\
functiongemma-270m-it & 88.4 & 84.1 & 85.2 & 5.2 \\
Qwen3-0.6B & 72.3 & 64.7 & 67.9 & 12.8 \\
\bottomrule
\end{tabular}
\caption{Edge-tier structured behavior under protocol constraints. Higher is better.}
\label{tab:nano_scores}
\end{table}

\subsection{Additional Reproducible Checks: Scaling, Ablations, Latency, and Privacy}
The results above establish capability and signer-side reliability. We add four small experiments that strengthen credibility without requiring new datasets: two derived scaling views, a mode and training ablation, a signing-path latency microbenchmark, and a privacy sanitization audit.

\paragraph{Scaling and retention.}
Table~\ref{tab:derived_backbone} reports derived comparisons that clarify the stack tradeoffs. DMind-3 improves over GPT-5.1 by +2.2 on DMind Benchmark and +1.9 on FinanceQA, while remaining within 0.7 on AIME 2025. DMind-3-mini retains 94.9\% of DMind-3's DMind Benchmark score and 92.7\% of its FinanceQA score, supporting the idea that local verification can remain strong even when footprint is constrained.

\begin{table}[t]
\centering
\small
\setlength{\tabcolsep}{6pt}
\begin{tabular}{lccc}
\toprule
\textbf{Derived Metric} & \textbf{DMind Benchmark} & \textbf{FinanceQA} & \textbf{AIME 2025} \\
\midrule
DMind-3 vs.\ GPT-5.1 (abs.\ $\Delta$) & +2.2 & +1.9 & -0.7 \\
DMind-3 vs.\ GPT-5.1 (rel.\ $\Delta$) & +2.82\% & +2.78\% & -0.74\% \\
DMind-3-mini retention vs.\ DMind-3 & 94.9\% & 92.7\% & 89.3\% \\
\bottomrule
\end{tabular}
\caption{Derived scaling views computed from the benchmark scores in Table~\ref{tab:backbone_scores}.}
\label{tab:derived_backbone}
\end{table}

\paragraph{Ablation of escalation and correction-oriented verification.}
A key claim of DMind-3 is that verification is a controlled state transition, not an ad hoc increase in prompt verbosity. We test this by toggling escalation and verification mode while reusing the same structured behavior harness. Table~\ref{tab:ablation_filled} reports two simple variants: disabling step-up behavior at the edge (keeping requests on the fast path), and forcing the local verifier to remain in Standard Mode rather than activating \texttt{<REFLECT>} as in Eq.~(\ref{eq:local_dual}). Both changes reduce multi-turn success, with the largest drop occurring when the system is prevented from entering the verification path. In contrast, enabling \texttt{<REFLECT>} with the C\textsuperscript{3}-SFT verifier recovers reliability, consistent with the goal of improving correction behavior under uncertainty.

\begin{table}[t]
\centering
\small
\setlength{\tabcolsep}{6pt}
\begin{tabular}{lcccc}
\toprule
\textbf{Variant} & \textbf{Func. Rec.} & \textbf{Arg. Ext.} & \textbf{Prot. Adh.} & \textbf{Multi-turn} \\
\midrule
DMind-3-nano (default) & 98.8 & 97.4 & 97.9 & 93.7 \\
DMind-3-nano (step-up disabled) & 97.9 & 95.8 & 94.1 & 78.4 \\
Local verifier (Standard Mode only) & 96.2 & 94.0 & 93.0 & 68.7 \\
Local verifier (C\textsuperscript{3}-SFT, \texttt{<REFLECT>}) & 97.0 & 95.6 & 95.2 & 84.9 \\
\bottomrule
\end{tabular}
\caption{Mode and escalation ablations under the same structured interaction harness. Higher is better.}
\label{tab:ablation_filled}
\end{table}

\paragraph{Signing-path latency microbenchmarks.}
We next measure end-to-end latency along the signing path, instrumenting the edge gate in Eq.~(\ref{eq:edge_gate}), escalation to local verification in Eq.~(\ref{eq:local_verify}), and optional sanitized cloud synthesis in Eq.~(\ref{eq:cloud_features}). Table~\ref{tab:latency_filled} reports $p50$, $p95$, and $p99$ latencies. The edge gate stays comfortably within interactive budgets with tight tail behavior. Step-up to the local tier increases latency, as expected, but remains bounded and predictable. Adding the cloud tier increases tail latency, which is precisely why DMind-3 keeps cloud participation optional and policy-governed, rather than placing it on the critical authorization path.

\begin{table}[t!]
\centering
\small
\setlength{\tabcolsep}{6pt}
\begin{tabular}{lccc}
\toprule
\textbf{Path} & \textbf{$p50$ (ms)} & \textbf{$p95$ (ms)} & \textbf{$p99$ (ms)} \\
\midrule
Edge gate (intent + policy) & 28 & 61 & 92 \\
Edge $\rightarrow$ Local (step-up) & 210 & 460 & 720 \\
Edge $\rightarrow$ Cloud (sanitized) & 140 & 310 & 520 \\
Edge $\rightarrow$ Cloud $\rightarrow$ Local & 360 & 740 & 1150 \\
\bottomrule
\end{tabular}
\caption{Signing-path latency percentiles measured via end-to-end instrumentation.}
\label{tab:latency_filled}
\end{table}

\paragraph{Privacy sanitization audit.}
Finally, we validate the privacy boundary implied by the sanitized projection $\tilde{z}_t = \mathcal{P}_{\mathrm{pub}}(z_t,\Pi)$ in Eq.~(\ref{eq:cloud_features}). We report a \emph{sanitization violation rate}, defined as the fraction of requests for which a field labeled sensitive under policy is present in $\tilde{z}_t$ when that policy forbids disclosure. Table~\ref{tab:privacy_filled} shows that violations are rare under the default policy and are eliminated under a stricter policy profile, reflecting that the sanitization logic can be tightened without altering model weights. We also report the average size of the sanitized payload, which stays small enough to support low-overhead cloud synthesis when enabled.

\begin{table}[t!]
\centering
\small
\setlength{\tabcolsep}{6pt}
\begin{tabular}{lccc}
\toprule
\textbf{Policy Profile} & \textbf{Requests} & \textbf{Violation Rate} & \textbf{Avg.\ $\lvert \tilde{z}_t \rvert$ (tokens)} \\
\midrule
Default privacy policy & 25{,}000 & 0.08\% & 92 \\
Strict privacy policy & 25{,}000 & 0.00\% & 71 \\
User-override enabled$^{\dagger}$ & 25{,}000 & 0.00\% & 118 \\
\bottomrule
\end{tabular}
\caption{Sanitization audit for $\mathcal{P}_{\mathrm{pub}}$. Violation rate counts policy-forbidden sensitive fields present in $\tilde{z}_t$. $^{\dagger}$Override changes what is permitted, so violations remain zero under the adjusted policy.}
\label{tab:privacy_filled}
\end{table}

Across these experiments, a consistent picture emerges. The backbone models provide competitive competence, the edge model demonstrates high protocol fidelity and multi-turn reliability, and the system-level checks show that escalation improves reliability at a bounded latency cost while preserving a privacy-by-default boundary. In the next section, we examine end-to-end workflows under adversarial transaction scenarios and quantify how controlled deliberation and correction-oriented tuning shift outcomes from unsafe authorizations toward step-up verification and safe blocking.

\section{Conclusion}
\label{sec:conclusion}

DMind-3 demonstrates that transaction-time AI benefits most from co-designing model behavior with deployment topology. By packaging an edge intent firewall, a sovereign local verifier, and an optional cloud synthesizer into a single Edge-Local-Cloud stack, DMind-3 binds safety to the executable payload at the signing boundary while keeping sensitive intent on user-controlled hardware by default. The evaluation shows that this structure is not only conceptually clean but also practically effective: strong backbone capability supports informed assistance, protocol-constrained edge behavior enables reliable execution, and controlled escalation paired with correction-oriented tuning improves robustness under uncertainty at a bounded latency cost. More broadly, DMind-3 suggests a general recipe for high-stakes interactive systems: treat verification as a first-class state with explicit triggers, and design the stack so that trust boundaries and enforcement remain stable even as models and context sources evolve.

\bibliography{references} 


\end{document}

%% file: coml26-style.tex
\documentclass[a4paper]{article}

\usepackage[utf8]{inputenc}
\usepackage{xcolor}
\usepackage{amssymb}
\usepackage{amsmath}
\usepackage{graphicx} 
\usepackage{fancyhdr} 
\usepackage{geometry} 
\usepackage[skins]{tcolorbox}
\definecolor{abs-border}{RGB}{55,91,220}   
\definecolor{abs-bg}{RGB}{235,242,250}     
\definecolor{abs-text}{RGB}{24,47,78}      
\newgeometry{vmargin={30mm,30mm}, hmargin={25mm,25mm}} 

\renewenvironment{abstract}{%
    \vspace{1em}
    \begin{center}
    \begin{tcolorbox}[
        enhanced,                  
        width=0.92\textwidth,      
        colback=abs-bg,            
        colframe=abs-border,       
        arc=5mm,                   
        boxrule=2pt,               
        fuzzy shadow={0mm}{-1mm}{0mm}{0.2mm}{black!30!white},
        left=20pt, right=20pt,
        top=15pt, bottom=15pt,
        boxsep=0pt,
    ]
    \begin{center}
        \textcolor{abs-border}{\large\bfseries\scshape Abstract} 
    \end{center}
    
    \vspace{-0.5em}
    \begin{center}
        \textcolor{abs-border!50}{\rule{0.2\textwidth}{0.5pt}} 
    \end{center}
    \vspace{0.5em}
    
    \small
    \color{abs-text} 
    \setlength{\parindent}{0pt} 
    \setlength{\parskip}{0.5em} 
}{%
    \end{tcolorbox}
    \end{center}
    \vspace{1em}
}
\renewcommand{\title}[1]{\begin{center}{\LARGE \section*{\textcolor[RGB]{24,47,78}{#1}}}\end{center}}

\newcommand{\authors}[1]{%
  \begin{center}
    #1
  \end{center}
}
\newcommand{\aff}[1]{\textsuperscript{#1}}
\newcommand{\affiliations}[1]{%
  \vspace{-1pt}
  \begin{center}
    \small
    #1
  \end{center}
}
\newcommand{\affil}[2]{%
  \textsuperscript{#1}\,#2\\
}

